\documentclass[10pt, draftcls, conference,onecolumn]{IEEEtran}

\IEEEoverridecommandlockouts
\usepackage{xspace,exscale,relsize}
\usepackage{fancybox,shadow}
\usepackage{graphicx}
\usepackage[usenames]{color}
\usepackage{amsfonts}
\usepackage{amssymb}
\usepackage{url}

\usepackage[noadjust]{cite}

\usepackage{amsmath}
	\makeatletter
	\let\over=\@@over \let\overwithdelims=\@@overwithdelims
	\let\atop=\@@atop \let\atopwithdelims=\@@atopwithdelims
  	\let\above=\@@above \let\abovewithdelims=\@@abovewithdelims
  	\makeatother
\usepackage{fixltx2e}

\usepackage{rotating}

\usepackage{ifpdf}

\usepackage{subfigure}
\usepackage{psfrag}

\newcommand{\matc}{\ensuremath{\mathcal{C}}}

\newcommand{\matx}{\ensuremath{\mathcal{X}}}

\newcommand{\matn}{\ensuremath{\mathcal{N}}}

\newcommand{\mreals}{\ensuremath{\mathbb{R}}}
\newcommand{\mcomp}{\ensuremath{\mathbb{C}}}

\ifx\eqref\undefined
	\newcommand{\eqref}[1]{~(\ref{#1})}
\fi
\ifx\mod\undefined
	\def\mod{\mathop{\rm mod}}
\fi

\newcommand{\vect}[1]{{\bf #1}}

\def\exp{\mathop{\rm exp}}

\DeclareMathOperator\EE{\rm\mathbb{E}}

\def\PP{\mathbb{P}}

\def\eqdef{\stackrel{\triangle}{=}}

\def\unifto{\mathop{{\mskip 3mu plus 2mu minus 1mu%
	\setbox0=\hbox{$\mathchar"3221$}%
	\raise.6ex\copy0\kern-\wd0%
	\lower0.5ex\hbox{$\mathchar"3221$}}\mskip 3mu plus 2mu minus 1mu}}

\ifx\lesssim\undefined
\def\simleq{{{\mskip 3mu plus 2mu minus 1mu%
	\setbox0=\hbox{$\mathchar"013C$}%
	\raise.2ex\copy0\kern-\wd0%
	\lower0.9ex\hbox{$\mathchar"0218$}}\mskip 3mu plus 2mu minus 1mu}}
\else
\def\simleq{\lesssim}
\fi

\ifx\gtrsim\undefined
\def\simgeq{{{\mskip 3mu plus 2mu minus 1mu%
	\setbox0=\hbox{$\mathchar"013E$}%
	\raise.2ex\copy0\kern-\wd0%
	\lower0.9ex\hbox{$\mathchar"0218$}}\mskip 3mu plus 2mu minus 1mu}}
\else
\def\simgeq{\gtrsim}
\fi

\newif\ifmapx
{\catcode`/=0 \catcode`\\=12/gdef/mkillslash\#1{#1}}
\edef\jobnametmp{\expandafter\string\csname papr_apx\endcsname}
\edef\jobnameapx{\expandafter\mkillslash\jobnametmp}
\edef\jobnameexpand{\jobname}
\ifx\jobnameexpand\jobnameapx
\mapxtrue
\else
\mapxfalse
\fi

\long\def\apxonly#1{\ifmapx{\color{blue}#1}\fi}

\newtheorem{remark}{Remark}
\newtheorem{theorem}{Theorem}

\newtheorem{proposition}[theorem]{Proposition}

\usepackage[
            CJKbookmarks=true,
            bookmarksnumbered=true,
            bookmarksopen=true,
            colorlinks=true,
            citecolor=red,
            linkcolor=blue,
            anchorcolor=red,
            urlcolor=blue,
            pdfauthor={Yury Polyanskiy, Yihong Wu}
            ]{hyperref}

\usepackage{prettyref}
\newrefformat{eq}{(\ref{#1})}
\newrefformat{chap}{Chapter~\ref{#1}}
\newrefformat{sec}{Section~\ref{#1}}
\newrefformat{algo}{Algorithm~\ref{#1}}
\newrefformat{fig}{Fig.~\ref{#1}}
\newrefformat{tab}{Table~\ref{#1}}
\newrefformat{rmk}{Remark~\ref{#1}}
\newrefformat{clm}{Claim~\ref{#1}}
\newrefformat{def}{Definition~\ref{#1}}
\newrefformat{cor}{Corollary~\ref{#1}}
\newrefformat{lmm}{Lemma~\ref{#1}}
\newrefformat{prop}{Proposition~\ref{#1}}
\newrefformat{app}{Appendix~\ref{#1}}
\newrefformat{ex}{Example~\ref{#1}}

\newcommand{\ie}{i.e.\xspace}

\newcommand{\TV}{\mathrm{TV}}

\newcommand{\Expect}{\mathbb{E}}
\newcommand{\expect}[1]{\mathbb{E} \left[#1\right]}
\newcommand{\Prob}{\mathbb{P}}
\newcommand{\prob}[1]{{ \mathbb{P}\left[ #1 \right] }}
\newcommand{\linf}[1]{\left\|{#1} \right\|_{\infty}}

\newcommand{\pth}[1]{\left( #1 \right)}

\newcommand{\sth}[1]{\left\{ #1 \right\}}

\newcommand{\indc}[1]{{\mathbf{1}_{\left\{{#1}\right\}}}}

\newcommand{\naturals}{{\mathbb{N}}}

\newcommand{\eexp}{{\rm e}}

\newcommand{\calN}{{\mathcal{N}}}

\begin{document}

\title{Peak-to-average power ratio of good codes for Gaussian channel}

\author{Yury~Polyanskiy 
        and~Yihong Wu 
\thanks{Y. Polyanskiy is with the Department of Electrical Engineering and Computer
Science, MIT, Cambridge, MA, 02139 USA, e-mail: {\ttfamily yp@mit.edu}. Y. Wu is with Department of Electrical and Computer Engineering, University of Illinois Urbana-Champaign, Urbana, IL 61801, USA, e-mail: {\ttfamily yihongwu@illinois.edu}.}%
\thanks{ The work was supported in part by the Center for Science of Information (CSoI), an NSF
Science and Technology Center, under Grant CCF-0939370 and NSF CAREER award Grant no. CCF-12-53205.}}
%

\maketitle

\begin{abstract}
Consider a problem of forward error-correction for the additive white Gaussian noise (AWGN) channel. For finite blocklength codes the backoff from the channel capacity is inversely proportional to the square root of the blocklength. In this paper it is shown that codes achieving this tradeoff must necessarily have peak-to-average power ratio (PAPR) proportional to logarithm of the blocklength. This is extended to codes approaching capacity slower, and to PAPR measured at the output of an OFDM modulator. As a by-product the convergence of (Smith's) amplitude-constrained AWGN capacity to Shannon's classical formula is characterized in the regime of large amplitudes. This converse-type result builds upon recent contributions in the study of empirical output distributions of good channel codes. 
\end{abstract}

\begin{IEEEkeywords} Shannon theory, channel coding, Gaussian channels, peak-to-average power ratio,
converse
\end{IEEEkeywords}

%
\IEEEpeerreviewmaketitle

\section{Introduction}

In the additive white Gaussian noise (AWGN) communication channel a (Nyquist-sampled) waveform $x^n = (x_1,\ldots, x_n) \in
\mreals^n$ experiences an additive degradation:
\begin{eqnarray} Y_j &=& x_j + Z_j\,, Z_j \sim \matn(0,1)\label{eq:awgn}
\end{eqnarray}
%
%
where $Y^n = (Y_1,\ldots,Y_n)$ represent a (Nyquist-sampled) received signal. 
An $(n, M, \epsilon, P)$ error-correcting code is a pair of maps $f:\{1,\ldots, M\} \to \mreals^n$ and $g: \mreals^n \to
\{1,\ldots, M\}$ such that
	$$ \PP[W \neq \hat W] \le \epsilon\,,$$
where $W\in\{1,\ldots,M\}$ is a uniformly distributed message, and
\begin{eqnarray} 
X^n &=& f(W)\\
\hat W &=& g(Y^n) = g(f(W) + Z^n)\,,
\end{eqnarray}
are the (encoded) channel input and the decoder's output, respectively. The channel input is required to satisfy the power constraint 
\begin{equation}\label{eq:powerconst}
	 \|X^n\|_2 \triangleq \left(\sum_{j=1}^n |X_j|^2\right)^{1\over2} \le \sqrt{nP}\,.
\end{equation}
The non-asymptotic fundamental limit of information transmission over the AWGN channel is given by
	$$ M^*(n, \epsilon, P) \triangleq \max\{M: \exists(n,M,\epsilon,P)\mbox{-code}\}\,.$$
It is known that~\cite{PPV08}\footnote{As usual, all logarithms $\log$ and exponents $\exp$ are taken to
an arbitrary fixed base, which also specifies the information units. $Q^{-1}$ is the inverse
of the standard $Q$-function:
\begin{equation} Q(x) = \int_x^\infty {e^{-y^2}\over\sqrt{2\pi}} dy\,.
\end{equation}}
\begin{equation}\label{eq:disp}
	 \log M^*(n, \epsilon, P) = nC(P)  - \sqrt{nV(P)} Q^{-1}(\epsilon) + O(\log n)\,, 
\end{equation}
where the capacity $C(P)$ and the dispersion $V(P)$ are given by
\begin{eqnarray} C(P) &=& {1\over 2} \log (1+P)\,,\label{eq:awgncap}\\
	   V(P) &=& {\log e \over 2} {P (P+2)\over (P+1)^2}\,.
\end{eqnarray}

The peak-to-average power ratio (PAPR) of a codeword $x^n$ is defined as
	$$ \text{PAPR}(x^n) \triangleq {\|x^n\|_\infty^2 \over {1\over n} \|x^n\|_2^2}\,,$$
where $\|x^n\|_\infty = \max_{j=1\ldots n} |x_j|$. This definition of PAPR corresponds to the case
when the actual continuous time waveform is produced from $x^n$ via pulse-shaping and
heterodyning: 
$$ s(t) = \sum_{j=1}^n x_j g(t-j)  \cdot \cos(f_c t)\,, $$
where $g(t)$ is a bounded pulse supported
on $[-1/2, 1/2]$ and $f_c$ is a carrier frequency. Alternatively, one could employ an (ideal) DAC 
followed by a low-pass filter. Such implementation is subject to peak regrowth due to
filtering: the maximal amplitude of the signal
may be attained in between Nyquist samples, and thus the PAPR observed by the high-power amplifier 
may be even larger.

In this paper we address the following question: What are the PAPR requirements of codes that attain
or come reasonably close to attaining the performance of the best possible codes~\eqref{eq:disp}?
In other words, we need to assess the penalty on $\log M^*$ introduced by imposing, in
addition to~\eqref{eq:powerconst}, an amplitude constraint:
\begin{equation}\label{eq:ampconst}
	 \|X^n\|_\infty \le A_n\,,
\end{equation}
where $A_n$ is a certain sequence. If $A_n$ is fixed, then even the capacity term in~\eqref{eq:disp}
changes according to a well-known result of Smith~\cite{smith}. Here, thus, we focus on the case of
growing $A_n$.

Previously, we have shown,~\cite[Theorem 6]{YP12-lpnorm} and~\cite{PV12-optcodes-extended}, that
very good codes for AWGN automatically satisfy $A_n = O(\sqrt{\log n})$. Namely, for any 
constant $\gamma>0$ there exists $\gamma'>0$ such that \textit{any} code with
\begin{equation}\label{eq:lognach}
	 \log M \ge nC - \sqrt{nV(P)} Q^{-1}(\epsilon) - \gamma \log n
\end{equation}
has at least $M\over 2$ codewords with
	$$ \|x^n\|_\infty \le \gamma'\sqrt{\log n}$$
In other words, very good codes cannot have PAPR worse than $O(\log n)$. 
On the other hand, for capacity-achieving input $X_*^n \sim \calN(0,P)$, classical results from extremal value theory shows that the peak amplitude behaves with high probability according to $\linf{X_*^n} = \sqrt{2 P \log n} + o_{\Prob}(1)$ \cite{DN70}. Therefore it is reasonable to expect that good codes must also have peak amplitude scaling as $\sqrt{2 \log n}$.
Indeed, in this paper we show that, 
even under much weaker assumptions on coding performance than~\eqref{eq:lognach}, 
the PAPR of at least half of the codewords must be $\Omega(\log n)$.

Interestingly, the $\log n$ behavior of PAPR has been recently observed for various communication
systems implementing orthogonal
frequency division multiplexing (OFDM) modulation. To describe these results we need to introduce
several notions. Given $x^n\in \mcomp^n$ the baseband OFDM (with $n$
subcarriers) signal $s_b(t)$ is given by
	$$ s_b(t) = {1\over\sqrt{n}}\sum_{k=0}^{n-1} x_k e^{2\pi i {k t\over n}}\,,$$
whereas the transmitted signal is
\begin{equation}\label{eq:ofdm}
	 s(t) = \mathrm{Re}\left(  e^{2\pi i f_c t}  s_b(t) \right )\,, \qquad 0 \le t < n
\end{equation}
where $f_c$ is the carrier frequency. For large $f_c$, we have that PAPR of $s(t)$ may be
approximated as~\cite[Chapter 5]{SL07}
\begin{equation}\label{eq:paprdef}
	 \text{OFDM-PAPR}(x^n) \triangleq {\max_{t\in[0,n]} |s(t)|^2 \over {1\over n} \int_0^n |s(t)|^2 dt} 
		\approx {\max_{t\in[0,n]} |s_b(t)|^2 \over {1\over n} \sum_{k=0}^{n-1}
|x_k|^2} \triangleq \text{PMEPR}(x^n)\,, 
\end{equation}
where the quantity on the right is known as the peak-to-mean envelope power
(PMEPR).\apxonly{\footnote{Apxonly 
footnote:
In practice, it seems most systems do not actually produce signal $s_b(t)$, because this would
require a \textit{direct-synthesis} of $n$ sine-cosines. Instead, first the values $A_\ell =
s_b({\ell\over L}), \ell = 1, \ldots, Ln$ are computed via FFT and some DSP goodness. Here $L$ is
the oversampling rate. Then, a signal with cyclic prefix of duration $\Delta$ is formed as follows:
$$ s_{b,1}(t) = \sum_{\ell = -Ln\Delta}^{Ln} A_{\ell \,\mathrm{mod}\, Ln} \cdot \delta(t-{\ell\over L}) $$
Then this signal is passed through a time-invariant filter $h$ to produce another signal 
$$ s_{b,2}(t) = s_{b,1}*h (t) $$
Filter $h$ is chosen so that a) spectrum supported on $(-2\pi L, 2\pi L)$, b) unit spectrum on
$(-2\pi, 2\pi)$ and c) $h$ has very fast decaying tails (summable). All of these together guarantee that
$s_{b,2}$ approximates the ideal signal $s_b$ very well, in particular the $\sup_t |s_{b,2}(t)|$
is within a constant multiple of $\sup_t |s_b(t)|$. (See private communication with G. Wunder for
more details).

Punchline: Even when $x^n \to s_b(t)$ is not done by direct-synthesis, $\text{PMEPR}(x^n)$ is
still a rather good approximation (within $1\pm \epsilon$) to the actual PAPR experienced by the
amplifier.
}}

Note that
values of $s_b(\cdot)$ at integer times simply represent the discrete Fourier transform (DFT) of
$x^n$. Thus PMEPR is always lower bounded by 
\begin{equation}\label{eq:pme_lb}
	\text{PMEPR}(x^n) \ge {\|F x^n\|_\infty^2\over {1\over n}\|x^n\|_2^2}\,,
\end{equation}
where $F$ is the $n\times n$ unitary DFT matrix
		$$ F_{k,\ell} = {1\over \sqrt{n}} e^{2\pi i {k\ell\over n}}\,.$$
In view of~\eqref{eq:pme_lb}, it is natural to also consider the case where the amplitude
constraint~\eqref{eq:ampconst} is replaced with
\begin{equation} \|U x^n\|_\infty \le A_n\,, \label{eq:ampconst_x} 
\end{equation}
where $U$ is some fixed orthogonal (or unitary) matrix. Note that for large $n$ there exist some
(``atypical'') $x\in \mcomp^n$ such that the lower
bound~\eqref{eq:pme_lb} is very non-tight~\cite[Chapter 4.1]{SL07}. Thus, the
constraint~\eqref{eq:ampconst_x} with $U=F$ is weaker than constraining inputs to those
with small $\text{OFDM-PAPR}(x^n)$. Nevertheless, it will be shown even with this relaxation
$A_n$ is required to be of order $\log n$. 

The question of constellations in $\mcomp^n$ with good minimum
distance properties and small OFDM-PAPR
has been addressed in~\cite{PT00}. In particular, it was shown in \cite[Theorems 7 -- 8]{PT00} that the (Euclidean) Gilbert-Varshamov
bound is achievable with codes whose OFDM-PAPR is $O(\log n)$ -- however, see Remark~\ref{rmk:gv_papr}
below. Furthermore, a converse result is established in \cite[Theorem 5]{PT00} which gives
 a lower bound on the PAPR of an arbitrary code in terms of its rate, blocklength and the minimum distance.
When $x^n \sim
\matn_c(0,P)^n$, the resulting distribution of OFDM-PAPR was analyzed
in~\cite{OI01}.
For so distributed $x^n$ as well as $x^n$ chosen uniformly on the sphere, 
OFDM-PAPR tightly concentrates around $\log n$, cf.~\cite[Chapter 6]{SL07}. Similarly, 
if the components of $x^n$ are
independently and equiprobably sampled from the $M$-QAM or $M$-PSK
constellations\apxonly{\footnote{This is for fixed $M$ and $n\to\infty$.}}
OFDM-PAPR again sharply peaks around $\log n$, cf.~\cite{LW06}.
If $x^n$ is an element
of a BPSK modulated BCH code, then again OFDM-PAPR is around $\log
n$ for most codewords~\cite{LW06,SL07}.

Thus, it seems that most good constellations have a large OFDM-PAPR of order $\log n$. Practically, this
is a significant detriment for the applications of OFDM. A lot of 
research effort has been focused on designing practical schemes for
\textit{PAPR reduction}. Key methods include amplitude clipping and filtering
\cite{LC98}, partial transmit sequence \cite{MH97}, selected mapping \cite{BFH96}, tone
reservation and injection \cite{Tellado00}, active constellation extension \cite{KJ03},
and others -- see comprehensive surveys~\cite{HL05,JW08}. In summary, all these techniques
take a base code and transform it so as to decrease the PAPR at the output of the OFDM
modulator. In all cases, transformation degrades performance of the code (either
probability of error, or rate). Therefore, a natural question is whether there exist
(yet to be discovered) techniques that reduce PAPR without sacrificing much of the
performance.

This paper answers the question in the negative: the $\Theta(\log n)$ PAPR is
unavoidable unless a severe penalty in rate is taken.

\section{Main results}

We start from a simple observation that achieving capacity (without
stronger requirements like \prettyref{eq:lognach}) is possible with arbitrarily slowly growing PAPR:


\begin{proposition}\label{th:cap} Let $A_n \to \infty$. Then for any $\epsilon\in(0,1)$ there exists a sequence of $(n, M_n, \epsilon, P)$ codes
satisfying~\eqref{eq:ampconst} such that
	$$ {1\over n} \log M_n \to C(P), \quad n \to \infty\,. $$
\end{proposition}
\begin{IEEEproof}
Indeed, as is well known, e.g.~\cite[Chapter 10]{cover}, selecting $M_n = \exp\{nC(P)+o(n)\}$ codewords with
i.i.d. Gaussian entries $X_j \sim \matn(0,P)$ results (with high probability) in a codebook that has
vanishing probability of error under maximum likelihood decoding. Let us now additionally remove all
codewords violating~\eqref{eq:ampconst}. This results in a codebook with a random number $M'_n \le
M_n$ of codewords. However, we have
\begin{eqnarray} 
	\EE[M'_n] &=& M_n \PP[ \|X^n\|_\infty \le A_n]\\
	     &=& M_n \left(1 - 2Q\left(A_n\over \sqrt{P}\right)\right)^n\\
	     &=& M_n \cdot \exp\{o(n)\} = \exp\{nC(P) + o(n)\}\,.\label{eq:art1}
\end{eqnarray}
The usual random coding argument then shows that there must exist a realization of the codebook 
that simultaneously has small probability of error and number of codewords no smaller than ${1\over
3} \EE[M'_n]$.
\end{IEEEproof}

\begin{remark}\label{rm:papr-cap}
Clearly, by applying $U^{-1}$ first and using the invariance of the distribution of
noise $Z^n$ to rotations we can also prove that there exist capacity-achieving codes satisfying
``post-rotation'' amplitude constraint~\eqref{eq:ampconst_x}. A more delicate question is whether
there exist good codes with small PMEPR (which approximates OFDM-PAPR). In that
regard,~\cite{OI01} and~\cite[Chapter 5.3]{SL07} show that if $X^n \sim \matc\matn(0, P I_n)$
we have 
\begin{equation}\label{eq:pme_ap}
	 \PP[\text{PMEPR}(X^n) \le A_n^2] \approx  e^{- \sqrt{\pi\over 3} n A_n 
e^{-A_n^2}}\,.
\end{equation}
Thus, repeating the expurgation argument in~\eqref{eq:art1} we can show that there exists
codes with arbitrarily slowly growing OFDM-PAPR and achieving capacity. Furthermore, there
exist codes achieving expansion in~\eqref{eq:disp} to within $O(\sqrt{n})$ terms with 
OFDM-PAPR of order $\log n$. 
\end{remark}

\begin{remark}\footnote{This result was obtained in collaboration with Dr. Yuval Peres
	\texttt{<peres@microsoft.com>}.} Not only capacity, but also the 
	Gilbert-Varshamov (GV) bound on the sphere in $\mreals^n$ can be achieved with
	arbitrarily slow growing PMEPR, that is, $A_n = \omega(1)$. Note that previously~\cite[Theorems 7 --
	8]{PT00} only showed the attainability of the GV bound with $A_n = \Theta(\sqrt{\log n})$.
	Indeed, since the GV bound follows from a greedy procedure, it is sufficient to
	show that for arbitrary $A_n \to \infty$ we have
	\begin{equation}\label{eq:rgp1}
			\PP[\text{PMEPR}(X^n) \le A_n^2] = e^{o(n)}\,, 
\end{equation}	
	where $X^n$ is uniformly distributed on a unit sphere $\mathbb{S}^{n-1}\subset\mreals^n$.
	Furthermore, we may take $X^n = Z^n / \|Z^n\|_2$ with $Z^n \sim \matn(0, I_n)$. Since
	$\|Z^n\|_2$ exponentially concentrates around $(1\pm\epsilon)\sqrt{n}$, statement~\eqref{eq:rgp1} is equivalent to
	\begin{equation}\label{eq:rgp2}
		\PP[\text{PMEPR}(Z^n) \le \mathrm{const}\cdot n A_n^2] = e^{o(n)}\,, 
	\end{equation}	
	Notice that for $Z^n$ being uniform on the hypercube $\{-1,+1\}^n$ the
	estimate~\eqref{eq:rgp2} was shown by Spencer~\cite[Section 5]{JS85}, and it implies
	achievability of the binary GV bound with $\omega(1)$ PMEPR -- see~\cite[Section
	5.4]{SL07}. From~\cite[(5.4)]{JS85} there exist vectors $L_j\in\mreals^n,
	j=1,\ldots,4n$ with norms $\|L_j\|_2=\sqrt{n}$ and such that~\eqref{eq:rgp2} is equivalent
	to 
	\begin{equation}\label{eq:rgp3}
			\PP\Big[\max_j |(L_j, Z^n)| \le \mathrm{const}\cdot \sqrt{n} A_n \Big] = e^{o(n)}\,. 
	\end{equation}
		Note that $\PP[(L_j, Z^n) \le \mathrm{const}\cdot \sqrt{n} A_n] = 1 - Q(A_n^{-1})  = e^{o(1)}$. Finally,~\eqref{eq:rgp3} follows from \v{S}id\'ak's lemma (see, e.g., \cite[(2.8)]{AG97}): 
		\[
		\PP\Big[\max_j |(L_j, Z^n)| \le \mathrm{const}\cdot \sqrt{n} A_n \Big] \geq (1 - Q(A_n^{-1}))^{4n} = e^{o(n)}.
		\]
	\label{rmk:gv_papr}
\end{remark}

\medskip

From Proposition~\ref{th:cap} it is evident that the question of minimal allowable PAPR is
only meaningful for good codes, i.e. ones that attain $\log M^*(n, \epsilon, P)$ to
within, say, terms of order $O(n^\alpha)$.  The following lower bound is the main result
of this note:
\begin{theorem}
	Consider an $(n, M, \epsilon, P)$-code for the AWGN channel with $\epsilon <
	1/2$
\begin{equation}\label{eq:exp1}
	 \log M \ge n C(P) - \gamma n^\alpha
\end{equation}
for some $\alpha\in[1/2, 1)$ and $\gamma > 0$. Define
	\begin{equation}
	\delta_{\alpha, P} = (1-\alpha) (\sqrt{1+P} - 1)^2.
	\label{eq:deltaap}
\end{equation}
Then for any $\delta < \delta_{\alpha, P}$, there exists
an $N_0=N_0(\alpha,P,\delta,\gamma,\epsilon)$, such that if $n \geq N_0$, then for any $n\times n$ orthogonal matrix
$U$ at least $\frac{M}{2}$ codewords satisfy
\begin{equation}\label{eq:exp2-U}
	 \|U x^n\|_\infty \ge \sqrt{2 \delta \log n}\,.
\end{equation}
\label{thm:An}	
\end{theorem}

\begin{remark}
The function $\alpha \mapsto \delta_{\alpha,P}$ suggests there exists a tradeoff between the convergence speed and the peak amplitude for a fixed average power budget $P$. Choosing $U$ to be the identity matrix, \prettyref{thm:An} implies that any sequence of codes with rate $C(P) - O(n^{-(1-\alpha)})$ needs to have PAPR at least
\[
\frac{2\delta_{\alpha,P}}{P} \log n = \frac{2(1-\alpha)(\sqrt{1+P}-1)^2}{P} \log n \, .
\]
In particular for $\alpha=\frac{1}{2}$, note that $\frac{\delta_{\frac{1}{2},P}}{P} \leq \frac{1}{2}$ for $P>0$. On the other hand, $X^n$ independently drawn from the optimal input distribution $\calN(0,P)$ has PAPR $2 \log n(1+o(1))$ with high probability regardless of $P$.
It is unclear what the optimal $\alpha$-$\delta$ tradeoff is or whether it depends on the average power $P$. 
\end{remark}

\begin{IEEEproof}
We start with a few simple reductions of the problem. 
	First, any code $\{\vect c_1,\ldots, \vect c_M\}\subset \mreals^n$ can be
	rotated to $\{U^{-1} \vect c_1, \ldots, U^{-1} \vect c_M\}$ without affecting the
	probability of error. Hence, it is enough to show~\eqref{eq:exp2-U} with $U=I_n$, the 
	$n\times n$ identity matrix. Second, by taking some $\epsilon'>\epsilon$ and reducing the 
	number of codewords from $M$ to $M'= c_\epsilon M$ we may further assume that the resulting
	$(n, M', \epsilon')$ subcode has small {\it maximal} probability of error, i.e. 
	$$ \PP[\hat W \neq i | W = i] \le \epsilon'\,, \qquad  i\in\{1,\ldots, M\}\,.$$ 
	Note that by Markov's inequality, $c_{\epsilon} \geq 1 - \frac{\epsilon}{\epsilon'}$. Since $\epsilon < 1/2$ we may have $c_\epsilon > 1/2$ by choosing $\epsilon' \in (2 \epsilon,1)$. 
	Third, if a resulting code contains less than $\frac{M}{2}$ codewords
	satisfying~\eqref{eq:exp2-U}, then by removing those codewords we obtain an $(n,
	M'', \epsilon', P)$ code such that 
	$$ \log M' \ge n C(P) - \gamma n^\alpha - \log \pth{c_\epsilon - {1\over2}} \triangleq nC(P) -
	\gamma' n^\alpha\,.$$
	Thus, overall by replacing $\gamma$ with $\gamma'$, $M$ with $M''$ and $\epsilon$ with $\epsilon'$ it is
	sufficient to prove: Any $(n, M, \epsilon, P)$ code with maximal probability of error
	$\epsilon$ satisfying~\eqref{eq:exp1} must have at least one codeword such that 
\begin{equation}\label{eq:exp2}
	 \|x^n\|_\infty \ge \sqrt{2 \delta \log n}\,,
\end{equation}
	provided $n\ge N_0$ for some $N_0 \in \naturals$ depending only on $(\alpha,\epsilon,P,\gamma,\delta)$. We proceed to showing the latter statement.

In~\cite[Theorem 7]{PV12-optcodes} (see also \cite{PV11-relent}) it was shown that for any 
$(n, M, \epsilon, P)$ code with maximal probability of error $\epsilon$ we have
	$$ D(P_{Y^n} || P_{Y^n}^*) \le nC(P) - \log M + a\sqrt{n}\,,$$
	where $a>0$ is some constant depending only on $(\epsilon,P)$, $P_{Y^n}^* =
\matn(0, 1+P)^n$ and $P_{Y^n}$ is the distribution induced at the output of the
channel~\eqref{eq:awgn} by the uniform message $W\in\{1,\ldots,M\}$. In the conditions of the theorem we have then
\begin{equation}\label{eq:exp3}
	 D(P_{Y^n} || P_{Y^n}^*) \le \gamma n^\alpha + a \sqrt{n} \leq \gamma' n^\alpha,
\end{equation}
where $\gamma'$ can be chosen to be $\gamma+a$.

Next we lower bound $D(P_{Y^n} || P_{Y^n}^*)$ by solving the following $I$-projection problem:
\begin{equation}\label{eq:iproj}
	 u_n(A) = \inf_{P_{Y^n}} D(P_{Y^n} || \matn(0,1+P)^n)\,,
\end{equation}
where $P_{Y^n}$ ranges over the following convex set of distributions:
	$$ P_{Y^n} = P_{X^n} * \matn(0,1)^n\,, \quad P_{X^n}[\|X^n\|_\infty \le A] = 1.
$$
Since the reference measure in~\eqref{eq:iproj} is of product type and $D(P_{U^n} || \prod_{i=1}^n Q_{U_i}) \geq \sum_{i=1}^n D(P_{U_i} || Q_{U_i})$, we have
\begin{equation}
u_n(A) = n u_1(A)\,. 	
	\label{eq:una}
\end{equation}

To lower bound $u_1(A)$, we use the Pinsker inequality \cite[p. 58]{ckbook}
\begin{equation}
D(P||Q) \geq 2\log e \, \TV^2(P,Q),
	\label{eq:pinsker}
\end{equation}
where the total variation distance is defined by $\TV(P,Q) = \sup_E |P(E)-Q(E)|$ with $E$ ranging over all Borel sets. Next we lower bound $\TV(P_{Y_1}, \calN(0,1+P))$ in a similar manner as in \cite[Section VI-B]{WV11-finconst}. To this end, let $Y_1^* \sim \calN(0,1+P)$. Fix $r > \frac{1}{\sqrt{1+P} - 1}$. Since $\prob{|X_1|\leq A}=1$, applying union bound yields
\begin{align}
     \prob{|Y_1| > r \sqrt{1+P} A } 
\leq & ~ \prob{|Z_1| > A (r \sqrt{1+P}-1)} \nonumber\\
= & ~ 2 Q(A (r \sqrt{1+P}-1)). \label{eq:Ym.K}
\end{align}
On the other hand, 
    \begin{equation}
    \prob{|Y_1^*| >  r \sqrt{1+P} A} = 2 Q(r A).
    \label{eq:Yinf.K}
\end{equation}
Assembling \prettyref{eq:Ym.K} and \prettyref{eq:Yinf.K} gives
\begin{align}
\TV(P_{Y_1}, \calN(0,1+P)) 
\geq & ~ Q(r A)  - 2 Q(A (r \sqrt{1+P}-1)) \label{eq:tv}.
\end{align}
Combining \prettyref{eq:pinsker} and \prettyref{eq:tv}, we have
\begin{eqnarray} 
u_1(A) &\ge& \left( Q(rA) - Q( (r\sqrt{1+P} -1 ) A) \right)^2  8\log e\,. 
		  \label{eq:iproj2}
\end{eqnarray}

Suppose that $A_n  \triangleq \linf{X^n} \leq \sqrt{2 \delta \log n}$. Let $r = \frac{1}{\sqrt{1+P} -1} - \tau$ with $\tau > 0$. 
Note that for all $x>0$,
\begin{equation}
\frac{x \varphi(x)}{1+x^2} \leq Q(x) \leq \frac{\varphi(x)}{x}	
	\label{eq:mills}
\end{equation}
 where $\varphi(x) = {1\over \sqrt{2\pi}} e^{-x^2/2}$ is the standard normal density.
Assembling \prettyref{eq:exp3}, \prettyref{eq:iproj}, \prettyref{eq:una} and \prettyref{eq:iproj2}, we have
\begin{align}
\gamma' n^{\alpha-1}
\geq & ~ \left( Q(r \sqrt{2 \delta \log n}) - Q( (r\sqrt{1+P} -1 ) \sqrt{2 \delta \log n}) \right)^2  8\log e	\nonumber \\
\geq & ~ c_1 \frac{n^{- \delta r^2}}{\sqrt{\log n}}.
\end{align}
for all $n \geq N_0$, where $c_1$ and $N_0$ only depend on $P$ and $\tau$. Hence
\[
\delta \geq \frac{1-\alpha-\frac{c_2 \log \log n}{\log n}}{r^2}.
\]
for some constant $c_2$ only depends on $P$ and $\tau$. By the arbitrariness of $\tau$, we complete the proof of \prettyref{eq:exp2}.
\end{IEEEproof}

\begin{theorem} Any $(n, M, \epsilon, P)$ code with maximal probability of error
$\epsilon$ must contain a codeword $x^n$ such that
	\begin{equation}
	 \|x^n\|_\infty \ge A
	\label{eq:peak-nonasym}
\end{equation}	
	where $A$ is determined as the solution to
	$$ \left( Q(r^* A) - Q( (r^* \sqrt{1+P} -1  ) A) \right)^2  8\log e = C - {1\over n} \log
M +\sqrt{6(3+4P)\over n} \log e + {1\over n}\log{2\over 1-\epsilon}\,,$$
	where 
\begin{equation}\label{eq:pno_r}
	 r^* = \frac{\sqrt{A^2+P \log (P+1)}+A \sqrt{P+1}}{A P}\,. 
\end{equation}
\end{theorem}
\begin{remark}[Numerical evaluation] Consider SNR=20~dB ($P=100$),
$\epsilon=10^{-3}$ and blocklength $n=10^4$. Then, any code achieving $95\%$, $99\%$
and $99.9\%$ of the capacity is required to have PAPR $-1.2$~dB (trivial bound), $1.99$~dB
and $3.85$~dB, respectively.
\end{remark}
\begin{IEEEproof}
The proof in~\cite{PV12-optcodes} actually shows
	$$ D(P_{Y^n} || P_{Y^n}^*) \le nC - \log M + \sqrt{6n(3+4P)} \log e + \log{2\over 1-\epsilon} \,. $$
	Let $A_n = \|x^n\|_\infty$.
	Using $D(P_{Y^n} || P_{Y^n}^*) \geq n u_1(A_n)$ and the lower bound on $u_1(A)$ in
\prettyref{eq:iproj2}, we obtain the result after noticing that the right-hand side
of~\eqref{eq:iproj2} is maximized by choosing $r$ as in~\eqref{eq:pno_r}.
\end{IEEEproof}

\section{Amplitude-constrained AWGN capacity}
	\label{sec:smith}

As an aside of the result in the previous section, we investigate the following question: How fast does the amplitude-constrained AWGN capacity converges to the classical AWGN capacity when the amplitude constraint grows? 
To this end, let us define
\begin{equation}
C(A,P) = \sup_{\substack{\expect{X^2} \leq  P \\ |X| \leq A \text{ a.s.}}} I(X; X+Z)
	\label{eq:CAP}
\end{equation}
This quantity was first studied by Smith \cite{smith}, who proved the following: For all $A,P>0$, $C(A,P) < C(\infty,P)=\frac{1}{2} \log(1+P)$. Moreover, the maximizer of \prettyref{eq:CAP}, being clearly non-Gaussian, is in fact finitely supported. Little is known about the cardinality or the peak amplitude of the optimal input. 
Algorithmic progress has been made in \cite{huang.meyn.2005} where an iterative procedure for computing the capacity-achieving input distribution for \prettyref{eq:CAP} based on cutting-plane methods is proposed. 
On the other hand, the lower semi-continuity of mutual information immediately implies that $C(A,P) \to \frac{1}{2} \log(1+P)$ as $A\to \infty$. A natural ensuing question is the speed of convergence. The next result shows that the backoff to Gaussian capacity due to amplitude constraint vanishes at the same speed as the Gaussian tail.

\begin{theorem}\label{thm:smith}
For any $P>0$ and $A\to\infty$ we have
\begin{equation}
e^{- \frac{A^2}{(\sqrt{1+P}-1)^2} + O(\ln A)} \leq \frac{1}{2} \log(1+P) - C(A,P)
\leq e^{- \frac{A^2}{2 P} + O(\ln A)}.
	\label{eq:smith-exponent}
\end{equation}
\end{theorem}

\begin{remark} 
Non-asymptotically, for any $A, P > 0$, the lower (converse) bound in~\eqref{eq:smith-exponent} is
\begin{equation}\label{eq:smith-conv}
	 \pth{\frac{ (\sqrt{1+P}-1) \log (1+P) }{A + A_1}}^2
\varphi^2\pth{\sqrt{1+P} {A_1\over P} + {A\over P}} 8\log e\,, 
\end{equation}
	and the upper (achievability) bound is
\begin{equation}\label{eq:smith-ach}
\frac{1}{1 - 2 Q(\theta)} \sth{  Q(\theta) \log\pth{1 + \frac{A \sqrt{P}}{1+P}\cdot \frac{\varphi(\theta)}{Q(\theta)}  }  + h(2 Q(\theta)) }
\end{equation}
	where $\theta \triangleq \frac{A}{\sqrt{P}}$, $A_1 \triangleq  \sqrt{A^2 + P \log (1+P)}$, and $h(\cdot)$ denotes the binary entropy function.
	\label{rmk:smith-nona}
\end{remark}

\begin{remark}
\prettyref{thm:smith} focuses on the fixed-$P$-large-$A$ regime where the achievability is done by choosing a truncated Gaussian distribution as the input. It is interesting to compare our results to the case where $A$ and $\sqrt{P}$ grow proportionally. To this end, fix $\alpha > 1$ and let $A = \sqrt{\alpha P}$. It is proved in \cite[Theorem 1]{KMF12} that as  $P \to \infty$, $\frac{1}{2} \log(1+P) - C(\sqrt{\alpha P},\sqrt{P}) \to L(\alpha)$, where $L(\alpha)$ can be determined explicitly \cite[Eq. (21)]{KMF12}. 
Moreover, let us denote the capacity-achieving input for \prettyref{eq:CAP} by $X_{A,P}^*$. Then as $P \to \infty$, $\frac{1}{\sqrt{P}} X_{\sqrt{\alpha P},P}^*$ converges in distribution to the uniform distribution (resp. a truncated Gaussian distribution) on $[-\sqrt{\alpha},\sqrt{\alpha}]$ if $\alpha \leq 3$ (resp. $\alpha > 3$). 
In particular, $L(3)  = \frac{1}{2} \log \frac{\pi \eexp}{6}$ corresponds to the classical result of 1.53dB shaping loss \cite{forney.ungerboeck.survey}.
The non-asymptotic bounds in \prettyref{rmk:smith-nona} yields a suboptimal estimate to $L(\alpha)$ in the proportional-growth regime.
\apxonly{Therefore this will not recover the shaping gain result.}
\end{remark}

%

\begin{IEEEproof}
	The lower bound follows from the proof of \prettyref{thm:An} by noting that for any $X$ such that $\Expect[X]=0$, $\Expect[X^2] \leq P$ and $|X|\leq A$, 
	\begin{align}
	\frac{1}{2} \log(1+P) - I(X; X + Z)
\geq & ~ \frac{1}{2} \log(1+ \Expect[X^2]) - I(X; X + Z)	\nonumber \\
= & ~ D(P_{X+Z} \, || \, \calN(0,1+\Expect[X^2]))	\nonumber \\
\geq & ~ D(P_{X+Z}\,  || \, \calN(0,1+P))	\label{eq:ngo1} \\
\geq & ~ u_1(A) \label{eq:ngo2} \\
\geq & ~ \left( Q(r^* A) - Q( (r^* \sqrt{1+P} -1  ) A) \right)^2  8\log e, \label{eq:ngo3} 
\end{align}
where \prettyref{eq:ngo1} follows from the fact that $\inf_{s > 0} D(P_{Y} \, || \, \calN(0,s)) = D(P_{Y} \, || \, \calN(0,\Expect[Y^2]))$ for all zero-mean $Y$, while \prettyref{eq:ngo2} and \prettyref{eq:ngo3} follow from \prettyref{eq:iproj} and \prettyref{eq:iproj2} with $r=r^*$ as in \prettyref{eq:pno_r}, respectively. 
We can then further lower bound \prettyref{eq:ngo3} by $8 \log e \varphi^2(b) (b-a)^2$, where 
$$b \triangleq \sqrt{1+P} {A_1\over P} + {A\over P} > 
 a \triangleq  \sqrt{1+P} {A\over P} + {A_1\over P} $$
The proof of \prettyref{eq:smith-conv} is completed upon noticing that 
$$ b-a = \frac{(\sqrt{1+P}-1) \log(1+P) }{A + A_1} \,.$$

To prove the upper bound, we use the following input distribution: Let $X_* \sim
\calN(0,P)$. Let $X_A$ and $\bar{X}_A$ be distributed according to $X_*$ conditioned on
the event $|X_*| \leq A$ and $|X_*| > A$, \ie, $\prob{X_A \in \cdot} = \frac{\prob{X_* \in
\cdot \cap [-A,A]}}{\prob{X_* \in [-A,A]}}$. Then in view of \prettyref{eq:mills} we have
\begin{align}\expect{X_A^2} &= P - \frac{2\theta P \varphi(\theta) }{1 - 2 Q(\theta)} < P\\
\Expect[\bar{X}_A^2] &= P + \frac{\theta P \varphi(\theta) }{Q(\theta)}. 
\label{eq:XAbar-power}
\end{align}
Then
\begin{align*}
 \frac{1}{2} \log(1+P)
= & ~ I(X_*; X_* + Z) \\
= & ~ I(X_*, \indc{|X_*| > A}; X_* + Z)	\\
\leq & ~ I(X_A; X_A + Z) \prob{|X_*| \leq A} + I(\bar{X}_A; \bar{X}_A + Z) \prob{|X_*| > A} + H(\indc{|X_*| > A}).
\end{align*}
In view of \prettyref{eq:XAbar-power}, we have
\[
(1-2Q(\theta)) I(X_A; X_A + Z) \geq \frac{1}{2} \log(1+P) - Q(\theta) \log\pth{1+ P +
A\sqrt{P} \frac{\varphi(\theta)}{Q(\theta)}  } - h(2 Q(\theta)) ,
\]
completing the proof of \prettyref{eq:smith-ach}.
\end{IEEEproof}

\section*{Acknowledgment}
The authors thank Dr. T. Koch for stimulating discussions and for pointing out the reference \cite{KMF12}.

\ifmapx
\section{Remark on the application to DMC}

We wanted to close the discussion by demonstrating that the method employed for the Gaussian
channel yields correct-order estimates for the discrete memoryless channels (DMC) too. 

Let $P_{Y|X}$ be a DMC and consider an arbitrary cost function $c:\matx \to \mreals$.
Define 
\[
c^* = \inf\{\expect{c(X)}: P_X \text{ is capacity achieving}\}
\]
to be the minimal average cost among all capacity-achieving input distributions. We claim:
For any $\epsilon$ there is a constant $a_\epsilon$ such that any $(n, M, \epsilon)$
with 
\begin{equation}\label{eq:dsize}
	 \log M \ge nC - n^\alpha 
\end{equation}
contains a codeword $x^n$ with
\begin{equation}\label{eq:dcost}
	 {1\over n} \sum_{j=1}^n c(x_j) \ge c^* - a_\epsilon n^{\alpha-1\over 2}\,. 
\end{equation}

An easy way to see this is by reducing to a subcode with constant composition $P_n$.
Then~\eqref{eq:dsize} implies
	$$ I(P_n, P_{Y|X}) \ge C - O(n^{\alpha -1})\,.$$
On the other hand, being smooth, mutual information approaches its maximum (the capacity)
quadratically. Thus, for some capacity-achieving $P_X^*$ we must have
	$$ \TV(P_n,P_X^*) = O (n^{\alpha-1\over 2})\,.$$
Then, computing the $P_n$-average cost one obtains~\eqref{eq:dcost}.

Next, we demonstrate how the same estimate can be obtained via the general (non-DMC
specific) method we proposed above. Suppose an $(n, M, \epsilon)$ code is given such that every codeword $x^n$ satisfies 
\begin{equation}\label{eq:dmc_0}
	 {1\over n} \sum_{j=1}^n c(x_j) \le c^* - \delta\,.
\end{equation}
According to~\cite[Theorem 5]{PV12-optcodes} (under mild regularity conditions on $P_{Y|X}$) we have
\begin{equation}\label{eq:dmc_1}
	 D(P_{Y^n} || P_{Y^n}^*) \le nC(P) - \log M + a\sqrt{n}\,,
\end{equation}
	where $a>0$ is some constant depending only on $\epsilon$, $P_{Y^n}^*$ is the $n$-th power
of a capacity-achieving output distribution $P_Y^*$ and  $P_{Y^n}$ is the distribution induced at
the output of the DMC $P_{Y|X}^n$ by the uniform message $W\in\{1,\ldots,M\}$. On the other hand,
single-letterizing the $I$-projection we have
\begin{equation}\label{eq:dmc_iproj}
	 D(P_{Y^n} || P_{Y^n}^*) \ge n \min_{P_Y} D(P_Y || P_Y^*)\,,
\end{equation}
where minimum is over all $P_Y$ inducible by $P_X \stackrel{P_{Y|X}}{\to} P_Y$ such that
		$$ \EE[c(X)] \le c^* -\delta\,. $$
Assuming for simplicity that there exists a capacity-achieving distribution $P_X^*$ positive on
every $x\in \matx$, then for every input distribution $P_X$ we have
	$$ I(X;Y) = D(P_{Y|X} || P_Y^* | P_X) - D(P_Y || P_Y^*) = C - D(P_Y||P_Y^*)\,,$$
where $C = \max_X I(X;Y)$ is the capacity. Therefore, the minimum in the right-hand side
of~\eqref{eq:dmc_iproj} equals $C-C(c^*-\delta)$, where
	$$ C(\sigma) \triangleq \max_{X: \EE[c(X)] \le \sigma} I(X;Y)$$
is the \textit{capacity-cost} function. Collecting the estimates together we get that any $(n, M,
\epsilon)$ code with codewords satisfying~\eqref{eq:dmc_1} should have
$$ \log M \le nC(c^* - \delta) + a\sqrt{n}\,.$$

\textbf{Oops, this is trivially derivable from standard Wolfowitz's strong converse!}

\smallskip
\textbf{Old details (Yihong):}

Consider BSC($\delta$) with $\delta \leq \frac{1}{2}$. We are interested in the weight of the good codes. Using the intuition of the capacity input distribution, which is equiprobable Bernoulli, we expect the weight of the codewords to behave similarly to the Binomial distribution $\mathrm{Binom}(n,\frac{1}{2})$, which is equal to $\frac{n}{2} + O_P(\sqrt{n})$. To this end, let the cost function be $c(x)=\indc{x=1}$. Then
\begin{align}
C_\delta(\epsilon)
= & ~ \max_{X: \prob{X=1}\leq \frac{1}{2}-\epsilon} I(X; X \oplus Z) \\
= & ~ h\pth{\frac{1}{2} + (2 \delta - 1)\epsilon}-h(\delta), 	\label{eq:bsc-skew}
\end{align}
attained by $\prob{X=1}=\frac{1}{2}-\epsilon$. Then the Taylor expansion of $\epsilon\mapsto C_\delta(\epsilon)$ near zero is
\[
C_\delta(\epsilon) = 1 - h(\delta) - \frac{2 (1 -2 \delta)^2 }{\log 2} \epsilon^2 + O(\epsilon^4).
\]
Similarly, the other direction...

For achievability, use Bernoulli($\frac{1}{2}+n^{\frac{\alpha-1}{2}}$).

\section{PAPR and Gilbert-Varshamov}

Continuing remark~\ref{rm:papr-cap}, we show an improvement of the result of
Paterson-Tarokh~\cite[Theorem 8]{PT00} (PMEPR equal to $8\log n$) and
Sharif-Hassibi~\cite[Theorem 6]{SH04} (PMEPR equal to $\log n$):
\begin{theorem} Fix $\delta\ge0$ and denote the Gilbert-Varshamov rate
	$$ R_{GV}(\delta) = -{1\over 2} \log \left(\delta(1-{\delta\over 4})\right)\,. $$ 
Then there exist $2^{nR_{GV} +o(n)}$ codewords $x^n\in\mreals^n$ with the pairwise Euclidean
distance at least $\sqrt{\delta n}$, constant energy $\|x^n\|_2 = \sqrt{n}$ and small PMEPR:
	$$ \text{PMEPR}(x^n) \le \ln \ln n\,, $$
see definition~\eqref{eq:paprdef}.
\end{theorem}
\begin{IEEEproof} Sketch: Assuming approximation~\eqref{eq:pme_ap}
holds for $X^n$ uniform on the sphere of radius $\sqrt{n}$ and setting
	$$ A_n^2 = \ln \ln n $$
we get
	$$ e^{-A_n^2} A_n = o(1)\,.$$
Thus, from~\eqref{eq:pme_ap} we have
	$$ \PP[\text{PMEPR}(X^n) \le A_n^2] \approx e^{-o(n)}\,.$$
In other words, the fraction of the surface area of a sphere with good PMEPR is
subexponetially large. Therefore, we can apply the usual Gilbert-Varshamov argument,
cf.~\cite[Section XI]{CS59},
restricted to the subset of the sphere
	$$\{x^n: \|x^n\|_2 = \sqrt{n}, \text{PMEPR}(x^n) \le A_n^2\}\,,$$
to obtain $2^{nR_{GV} + o(n)}$ codewords with required minimum distance.

\textbf{TODO:} To make this argument rigorous, read and apply bounds from~\cite[Chapter
6]{SL07}. Instead of~\eqref{eq:pme_ap} another approximation that yields PMEPR=$O(\log
\log n)$ is extreme-value based estimate~\cite[(5.64)]{SL07}, see also~\cite{OI01}. 

\textbf{Further progress:} Let $X^n$ be uniform on the set $\{x^n \in \mcomp^n: \|x^n\|_2 =
\sqrt{n}\}$, $U^n$ be iid $\mathcal{CN}(0,1)$ then we have
$$ \PP[\text{PMEPR}(X^n) \le 2\gamma] \ge \PP[\text{PMEPR}(U^n) \le \gamma] - e^{-n E}\,.$$
Thus it is sufficient to restrict to $U^n$. On the other hand, by~\cite[Chapter 4]{SL07}, we have
(two-oversampling)
$$ \text{PMEPR}(U^n) \le C_2 \max(\max_{j} | \tilde U_j|^2, \max_j |\tilde V_j|^2)\,,$$
where
\begin{align} 
	\tilde U_j &= {1\over \sqrt{n}} \sum_{k=0}^{n-1} U_j e^{2\pi i {jk\over n}}\,,\\
	\tilde V_j &={1\over \sqrt{n}} \sum_{k=0}^{n-1} U_j e^{2\pi i {(j+{1\over2})k\over n}}\,.
\end{align}
Note that $\tilde U_j$ are also iid Gaussians $\sim \mathcal{CN}(0,1)$, and that $\tilde V_j$ are
expressible in terms of $\tilde U_j$ as follows
\begin{align} 
   \tilde V_j &= (\phi * U)(j) = \sum_{k=0}^{n-1} U_{j-k} \phi(k)\,,\\
   \phi(k) &= {1\over n} {i e^{-{i\pi\over 2n}(2k + 1)} \over \sin({\pi \over 2n}(2k+1)) } = 
   		{1\over n} + i \cdot \mathrm{ctg}{(2k+1)\pi\over 2n}
\end{align}
Hence, on one hand we have
$$ \PP[\max_{j} | \tilde U_j|^2 \le \gamma] = (1-2Q(\gamma))^n\,,$$
on the other hand we have
$$ \PP[\max_j |\tilde V_j|^2 \le \gamma \,|\, \max_{j} | \tilde U_j|^2 \le \gamma] =
	\PP[ \| W * \phi\|_\infty^2 \le \gamma]\,,$$
where convolution is over $\mathbb{Z}/n\mathbb{Z}$ and $W_j$ are iid truncated Gaussians:
$$ P[W_j \in (\cdot)] = \PP\bigg[ U_j \in (\cdot) \,\,\bigg|\,\, |U_j|^2 \le \gamma\bigg]\,. $$
Thus, sufficient to prove that whenever $\gamma = c \log \log n$ we have
$$ \PP[ \| W* \phi\|_\infty^2 \le \gamma] \ge e^{-o(n)}\,.$$

\textbf{Ideas:} Young inequality does not appear to work (but verify again, also applying
Marcinkewicz-Zygmund inequalities). However, I want to try
rearrangement lemmas which lead to order statistics estimates. Also check G. Peshkir's
paper (1995): ``Best constants in Kahane-Khintchine inequalities for complex Steinhaus
functions''.

\end{IEEEproof}

\textbf{Remarks (Aug. 2013)}:
\begin{enumerate}
	\item I was silly in proposing $A_n^2 = \ln \ln n$. My intuition was based purely
	on~\eqref{eq:pme_ap} and thus any $A_n \to \infty$ should also lead to
	subexponential rate backoff.
	\item Overall, the problem may be formulated as follows. Define two sequences of
	convex bodies:
	\begin{align} K_n &= \left\{x^n \in \mcomp^n: \left| \sum_{k=0}^{n-1} x_k e^{i\omega k}
		\right| \le 1\right\}\\
	   \tilde K_n &= \left\{x^n \in \mcomp^n: \|x\|_\infty \le 1, \|\Phi 
	   x\|_\infty \le 1 \right\}\,, 
   \end{align}
   	where 
	$\Phi = \{\phi(\ell-m)\}_{\ell, m}$ is a \textit{unitary} matrix (with
	spectrum $e^{\pi k/n}$, $k=0,\ldots,n-1$). 
	It is known that for some constant $c$
		$$ c \tilde K_n \subset U^* K_n \subset \tilde K_n $$
	where  $U = \{e^{i{\ell m \over n}}/\sqrt{n}\}_{\ell, m} $ is the DFT matrix.
	\item Let us call the sequence of convex bodies $K_n$
	\textit{subexponentially-fat} if 
\begin{equation}\label{eq:conj1}
		A_n \to \infty \implies \PP[ \|Z^n\|_{K_n} \le A_n ] \ge e^{-o(n)}\,,
\end{equation}
where the $K$-norm of the vector is 
$$ \|x\|_{K} = \inf\{s: x \in sK\} $$

Note that~\eqref{eq:conj1} is equivalent to
\begin{align} \lim_{A\to\infty} E(A) &=0 \label{eq:conj2} \\
E(A) &\eqdef \limsup_{n\to\infty} -{1\over n} \log \PP[Z^n \in A \cdot K_n]
\end{align}
	Proof: notice that $f_n(A) = -{1\over n} \log \PP[Z^n \in A \cdot K_n]$ is
	monotonically decreasing and do some $\epsilon-\delta$ arguments.
\item In words:~\eqref{eq:conj2} means that when $K_n$ is scaled suff. large it consumes
arbitrarily small exponent of Gaussian measure. For example $K_n = \{x: \|x\|_\infty \le
1\}$ has this property, while $K_n = \{x: \|x\|_2 \le n^{1-\epsilon}\}$ does not.

\item 
Our question is: Is it true that $K_n$ (equivalently $\tilde K_n$) -- is a sequence of
subexponentially-fat convex bodies?

\item More weakly: to improve previous results we only need to prove~\eqref{eq:conj1} for
any $A_n \ll \sqrt{\log n}$.

\item After some investigations I realized this question plays rather poorly against standard
tools of theory of maxima of Gaussian processes, concentration of measure etc. Basic
reason is:~\eqref{eq:conj1} asks a question about the very tail of the distribution of
$K$-norm of Gaussian $Z^n$. Most concentration results talk about ``the bulk''.

\item Also notice that for studying $\tilde K_n$ function $\phi_n(\cdot)$ plays an
very special role. It has many properties, but one of the important ones is
$\|\phi_n\|_{\ell_2}=1$. I could not even answer whether there is \textit{any} sequence
$\phi_n$ with unit $\ell_2$-norm such that~\eqref{eq:conj1}-\eqref{eq:conj2} would be
violated (\textbf{TODO:} random coding with $\phi$ having $\matc\matn(0,1/n)$ entries).
\end{enumerate}

\fi 



\end{document}